# PATTERN DETECTION WITH RARE ITEM-SET MINING

Mehdi Adda[1], Lei Wu[2], Sharon White[2], Yi Feng [3]

[1]Computer Science, University of Quebec at Rimouski, Rimouski, Canada
`mehdi_adda@uqar.qc.ca`
[2]Software Engineering, University of Houston-Clear Lake, Houston, USA
`wul,whites@uhcl.edu`
[3]Computer Science, Algoma University, Sault Ste. Marie, Canada
`feng@algomau.ca`

## ABSTRACT

*The discovery of new and interesting patterns in large datasets, known as data mining, draws more and more interest as the quantities of available data are exploding. Data mining techniques may be applied to different domains and fields such as computer science, health sector, insurances, homeland security, banking and finance, etc. In this paper we are interested by the discovery of a specific category of patterns, known as rare and non-present patterns. We present a novel approach towards the discovery of non-present patterns using rare item-set mining.*

## 1. INTRODUCTION

Data mining techniques are used to process huge amounts of information in order to extract hidden knowledge to be directly interpreted or exploited to feed other processes. In many cases, data mining techniques are used to discover patterns that can be of interest to a specific domain of application. A pattern is a collection of events/features that occur together in a transaction database. As a matter of fact, different categories of patterns exist, such as sequences, item-sets, association rules and graph patterns, etc. The choice of a technique depends not only on the nature of input data but also depends on what we want to obtain and what view or part of the data we want to be described or represented in a more intelligible and concise way. To filter out the patterns, some criteria are used. The most known criteria are the support and the confidence. While the support represents the number of times a pattern occurs in the initial database (a.k.a. frequency), the confidence represents a proportion value that shows how frequently a part of the pattern, called premise, occurs among all the records containing the whole rule body.

Herein, we classify pattern categories according to the specific use of the support threshold. In fact, by setting ranges for the support, one can obtain different categories of patterns. For example, if we set a minimum support threshold that a pattern has to satisfy to be considered as an interesting pattern, we obtain what is called frequent pattern. By setting a maximum support threshold we obtain another category of patterns called rare patterns. Whereas frequent patterns focus on mining patterns that appear more frequently in the database, rare patterns aim at discovering patterns that are less frequent.

Rare patterns can be used in different domains such as biology, medicine and security [9] [18][15], etc.. For example, by analysing clinical databases one can discover rare patterns that will help doctors to make decisions about the clinical care. In the security field, normal behaviour is very frequent, whereas abnormal or suspicious behaviour is less frequent. Considering a database where the behaviour of people in sensitive places such as airports are recorded, if we model those behaviours, it is likely to find that normal behaviours can be represented by frequent patterns and suspicious behaviours by rare patterns.

As one can notice, each category of patterns explores the data seeking for a specific kind of knowledge. It is noteworthy that other categories of patterns different from frequent and rare patterns can be mined. For example, one can be interested by the patterns with a frequency smaller that an upper limit and greater than a lower limit. Those categories of patterns may be used to answer questions such as what items/objects/products/events are visited/bought/occurred together with a minimum and/or maximum frequency. However, some questions may not be directly answered by the mentioned categories of patterns. For instance, with frequent and rare patterns someone may not find the collections of items/objects/products/events that are not visited/bought/occurred together. Answering such questions may be of interest in cases where we are more interested in what is missing than what is already existing, or what we already know about the existing data. Such patterns, we call here as non-present patterns, may potentially be used to detect what is missing in a defective process/situation and correct it by supplying us with candidate solutions that have not yet been encountered.

In this paper, we present a framework to represent different categories of interesting patterns and then instantiate it to the specific cases of rare and non-present patterns. Later on, we present a generic framework to mine patterns based on the Apriori approach [2]. The resulting approach is Apriori-like and the mining idea behind it is that if the item-set lattice representing the item-set space in classical Apriori approaches is traversed in a bottom-up manner, equivalent properties to the Apriori exploration of frequent item-sets is provided to discover rare item-sets. This includes an anti-monotone property and a level-wise exploration of the item-set space.

This paper is organized as follows: we start by an overview of the problem of item-set mining and the classical Apriori approach used to mine frequent patterns. Then we investigate the other kind of knowledge that is not usually discovered using frequent item-set mining: rare item-sets and non-present item-sets. In order to mine rare and non-present item-sets, we methodologically extract the principles behind Apriori approach, and then generalize it in such a way it can not only be applied to mine frequent item-sets but also to mine other categories of patterns such as rare and non-present item-sets. Our goal is to apply the principle of the Apriori approach used for frequent item-set mining and apply it to rare and non-present item-set mining. We start by factorizing the key elements of Apriori approaches such as the traversal of the item-set space, the pruning principle, the combination of item-sets at a level to generate new item-set in the next level, and then we propose an Apriori generalized framework that abstracts those elements. With the Apriori generalized framework on hands, the instantiation of a specific framework to tackle the Apriori based mining approaches for rare and non-present item-sets becomes straightforward. The remainder of this paper is organized as follows. After presenting rare and non-present item-set mining problems in Section 2, we discuss related works in Section 3. Then, the framework to represent different categories of interesting patterns is presented in Section 4. In Section 5 we begin by presenting the main idea behind our approach and show how the approach is effective using an example. Then, we present the generalized Apriori framework which by means of instantiation lead to an Apriori-based rare and non-present item-set mining approaches and an algorithm called ARANIM (*Apriori for Rare And Non-present Item-set Mining*). Before concluding, we present in Section 6 an approach to discover suspicious behaviour in the context of web applications.

## 2. RARE AND NON-PRESENT ITEM-SET MINING

In this section we present an example of rare and non-present item-set mining. The input data is composed of a database of transactions, and each transaction is identified by an id and composed of a set of items. In the real world, a transaction may be seen as the basket bought by a customer during a determined period of time (day, week, month, etc.). Each basket is composed of a set of items that are purchased together. In Table 1 we represent an abstract database, denoted by D, where the alphabet letters are considered as items. Given the database of transactions such as presented in Table 1, our goal is to find two categories of sets of items, also called item-sets. The first category is composed of item-sets that are present in at most two transactions, and the second category is composed of item-sets that do not occur in any transaction and composed of a maximum number of items equal to the cardinality of the largest transaction. The number of times an item-set occurs in the database is called the item-set support. In our case the maximum support is equal to 3.

Table 1. Transaction database

| id | transaction |
|---|---|
| $t_1$ | {a, b c, d} |
| $t_2$ | {b, d} |
| $t_3$ | {a, b, c, e} |
| $t_4$ | {c, d, e} |
| $t_5$ | {a, b, c} |

The set of all item-sets that can be generated from the transaction database is presented in *Figure 1* by a diagram of the subset lattice for five items with the associated frequencies in the database. In the lattice each level is composed of item-sets having the same length. The top element in the lattice is the empty set.

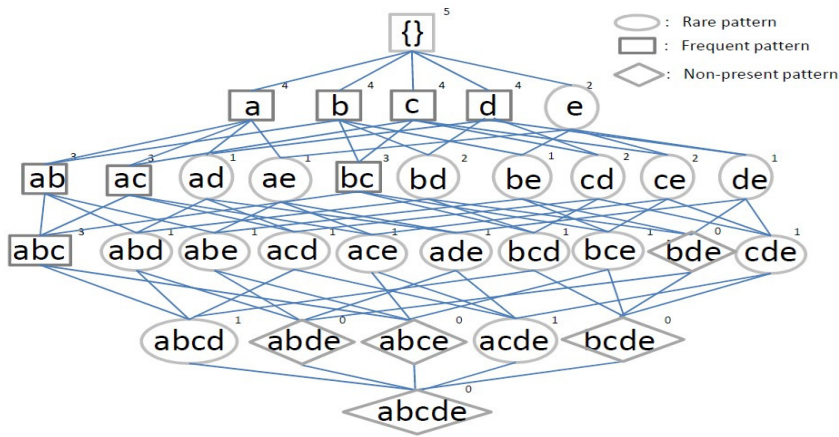

*Figure 1:* Lattice representing a hierarchically ordered space of item-sets and their frequencies. Frequent item-sets are square-shaped, rare item-sets are oval-shaped and non-present item-sets are diamond-shaped.

Each lower level k contains all of the item-sets of length k, also denoted k-item-sets and the last level contains an item-set composed of all items (i.e. *a, b, c, d, e*). Lines between nodes represent a subset relationship between item-sets. For each item-set we compute its support. For example, the item-set composed of the items b and d denoted by *<bd>* have 2 as support, and we denote it by *<bd,2>*. In fact, the item-set *bd* is present in the transactions $t_1$, $t_2$. The set of rare item-sets we are looking for are those in the lattice with support greater than 0 and less than 3. The rest of item-sets are either frequent (having support greater or equal to 3) or non-present (support=0).

In Figure 1 rare item-sets are drawn with ovals, frequent item-sets are drawn with rectangles where non-present item-sets are drawn with diamonds. Thus, after counting the frequencies of each item-set, we obtain the following set of rare item-sets <*e*,2>, <*ad*,1>, <*ae*,1>, <*bd*,2>, <*be*,1>, <*cd*,2>,<*cd*,2>, <*de*,1>, <*abd*,1>, <*abe*,1>, <*acd*,1>, <*ace*,1>, <*ade*,2>, <*bcd*,1>, <bce,1>, <*cde*,1>, <*abcd*,1>, <acde,1>. The set of non-present item-sets is composed of the following elements <abde,0>, <*abce*,0>, <*bcde*,0>, <abcde,0>.

## 3. RELATED WORK

The previous methods to mine item-sets can be divided into two categories, namely frequent item-set mining and rare item-set mining techniques. Whereas the problem of frequent item-set mining have been widely studied, the problem of rare item-set mining has just started to spark the researchers interest and non-present item-sets have not yet been addressed as independent mining task.

### 3.1. Frequent Item-set Mining

Item-set space has two properties, a monotone property and an anti-monotone property [12][2] [20]. Every non-empty subset of a frequent item-set is a frequent item-set, and every superset of a non-frequent item-set is non-frequent. Based on those properties, the algorithm Apriori [2] was developed to efficiently mine frequent item-sets. Later on, new algorithms have been proposed to mine frequent item-sets such as Eclat [24], FP-Growth [11] and TM [19]. A recent survey on frequent item-sets mining techniques is presented in [23].

The Apriori approach was successfully used to generate frequent item-sets contained in a transaction database [2] [3] [17]. Apriori approach exploits the monotonicity property of the support of item-sets. Apriori-based algorithms perform top-down breadth-first search through the space of all item-sets. In the first pass, the support of each individual item is counted, and the frequent ones are inserted to the frequent item-set set of level 1. In each subsequent pass, the frequent item-sets determined in the previous pass are used to generate new item-sets called candidate item-sets. The support of each candidate item-set is counted, and the frequent ones are determined. This process continues until no new frequent item-sets are found.

### 3.2. Rare and Non-Present Item-sets

Recently, a work presented in [21] proposes an approach to mine rare item-sets that is based on the Apriori algorithm used to mine frequent item-sets. The main idea consists at traversing the item-set space by the Apriori algorithm used to mine frequent item-sets and collect at each level the item-sets that are usually pruned out in the original algorithm and are used as seed for a second algorithm in order to mine the remaining rare item-sets. Another algorithm, called MINIT, is proposed in [10] to mine only minimal infrequent item-sets. A minimal infrequent item-set is an infrequent item-set that do not have a subset of items which forms an infrequent item-set. In other words, an infrequent item-set is minimal if all its proper subsets are frequent. It is noteworthy that the output of this algorithm may be used to mine all rare item-sets.

More recently, we proposed in [1] a framework that absorbs the spirit of the Apriori approach to tackle the problem of rare item-set mining. In this work, we started by factorizing the key elements of Apriori approaches such as the traversal of the item-set space, the pruning principle, the combination of item-sets at a level to generate new item-set in the next level, and then we proposed an Apriori generalized framework that abstracts those elements. However, in the proposed approach, no distinction is made between rare and non-present patterns. A performance comparison of our algorithm with the algorithm presented in [21] is also presented.

### 3.3. Intrusion Detection

Intrusion detection methods are of two types: anomaly detection and misuse (signature) detection. While anomaly detection techniques focus on the detection of user behaviour that is considered as abnormal [5] [22] [7], signature detection focuses on the identification of a behaviour that is similar to known cases that are considered as intrusions. A model is generally is used to represent the known intrusions [13] [16]. The main drawback of model based intrusion detection is the new attacks may be missed if not always kept up-to-date. In order to overcome this limitation, data mining and machine learning approaches are is the current trend to detect intrusion [14] [6] [4] [8] [25].

## 4. GENERALIZED FRAMEWORK FOR PATTERN MINING

### 4.1. Problem Definition

The general problem of pattern mining can be summarized as follows. Consider a universe of objects, or items, $I$, a universe $\Delta$ composed of sets of items from $I$, also called item-sets, a database $D$ made of records combining items from $I$ ($D \subseteq \Delta$), and a minimum frequency, or support, threshold $\delta_1$ (a.k.a *minsupp*), and a maximum frequency threshold $\delta_2$ (a.k.a *maxsupp*) (See *Definition 4.1*), such that $0 \leq \delta_1 < \delta_2 \leq |D|+1$ where $|D|$ is the number of records contained in $D$ also known as *cardinality*.

**Definition 4.1** *Let D be a subset of the data set* $\Delta$, $\Gamma$ *the pattern space, and f in* $\Gamma$, *the frequency, or support, of f, w.r.t. D, denoted by supp$_D$(f), corresponds to the number of records in D that are instances of f. More formally, we have the following:*
*supp$_D$:* $\Gamma \to \mathbb{N}$ *such that:* $\forall f \in \Gamma$: *supp$_D$(f)* $= \sum_{d \in \Gamma}(d \triangleleft f)$.

Two languages, the pattern language $\Delta$ composed of patterns and the data one $\Gamma$ composed of item-set transactions, and two binary relations underlay the problem: the *generality* between patterns, denoted by $\subseteq_p$, and *instantiation* between a data set record and a pattern, denoted by $\triangleleft$. When a transaction, say $d$, is instance of a pattern, $f$, $d \triangleleft f$ returns the value *1*, otherwise its value is *0*. Generality follows instantiation as given a pattern $f \in \Gamma$ and a super-pattern thereof $f'$ ($f \subseteq_p f'$), each record $d \in \Delta$ instantiating $f$ ($d \triangleleft f$) instantiates $f'$ as well.

We define the domain of interpretation of a pattern $f$ with respect to the data set space $\Delta$, denoted by $[f]_\Delta$, as the set of the pattern instances. When considering only a subset of the data set space, say $D$, each record in $D$ is instance of at least one pattern from $\Gamma$. The set of all those patterns, constitute what we call the *coverage set*. A formal definition is given below (see *Definition 4.2*).

**Definition 4.2** *Given a couple ($\Gamma$, $\Delta$) of pattern and data spaces, and the data set D such that* $D \subseteq \Delta$. *The coverage of the data set D w.r.t. the pattern space* $\Gamma$, *denoted by* $C_\Delta^\Gamma(D)$, *is a set of patterns such that each pattern has at least one instance in D. More formally, the mapping* $C_\Delta^\Gamma : 2^\Delta \to 2^\Delta$ *is defined by* $C_\Delta^\Gamma(D) = \{f \mid f \in \Gamma \text{ and } \exists d \in D \text{ such that } d \triangleleft f\}$.

**Remark 4.1** *The coverage set of the data set space is the pattern space ( $C_\Delta^\Gamma(\Delta) = \Gamma$ ).*

Pattern mining amounts to extracting the family $F_{\delta_1,\delta_2}^D$ of item-set collections, or patterns that are present in $D$ at least in $\delta_1$ records and at most in ($\delta_2 - 1$) records (see *Definition 3.3*). *Proposition 4.1* states that the coverage set is equivalent to the largest family of patterns.

**Definition 4.3** *Given $D$ from $\Delta$, and $\delta_1, \delta_2$ such that $0 \leq \delta_1 < \delta_2 \leq |D|+1$, the pattern family $F_{\delta_1,\delta_2}^D$ is a set of patterns such that $F_{\delta_1,\delta_2}^D = \{f \mid f \in C_\Delta^\Gamma(D) \wedge (\delta_1 \leq \sup p_D(f) < \delta_2)\}$. The patterns belonging to $F_{\delta_1,\delta_2}^D$ are also called **interesting patterns**.*

**Proposition 4.1** *Let $\Gamma, \Delta$ be the pattern and data spaces, respectively, $D$ be a subset of $\Delta$ and the family of patterns $F_{\delta_1,\delta_2}^D \subseteq \Gamma$, then $C_\Delta^\Gamma(D) = F_{\delta_1,\delta_2}^D$.*

In the remainder of this section, a pattern mining problem will be represented by two entities. The first component is the underling framework composed of the pattern space, the data space, the generalization and instantiation relationships. Thus, the generalized framework stood for the quadruplet $<\Gamma, \Delta, \subseteq_p, \triangleleft>$. The second component consists of the pattern family $F_{\delta_1,\delta_2}^D$ that represents a category of patterns contained in the data set $D$, subset of the data space, and that have support within the interval $[\delta_1, \delta_2[$. This family of patterns is also represented by the triplet $<D, \delta_1, \delta_2>$.

## 4.2. Frequent, Rare and Non-Present pattern mining instances of the generalized framework

Frequent, rare and non-present pattern mining are special cases of pattern mining and can be modelled using the framework defined above. The problem of frequent pattern mining consists at looking only for patterns with support at least equal to a fixed threshold. More formally, it is related to the extraction of the family of patterns $F_{\delta_1,\delta_2=|D|+1}^D$. However, in rare pattern mining we are exclusively looking for non-common patterns: patterns that are present in the data transactions with a frequency smaller than a fixed threshold. The problem is equivalent to extracting the family $F_{\delta_1=1,\delta_2}^D$ of patterns. In the case of non-present pattern mining, we are only interested by patterns that are not present and is related to the extraction of the family $F_{\delta_1=0,\delta_2=1}^D$ of patterns.

In all the three previous cases, both data set and pattern spaces are reduced to sets of item-sets, i.e., $\Delta = \Gamma = 2^I$, while $\subseteq_p$ and $\triangleleft$ boil down to set-theoretic inclusions. Hence the mining goal amounts to finding all the subsets of a family of patterns contained in $D \subseteq \Delta$. In other words, in the case where the data and pattern spaces are composed of items without generalization among them, the frequent, rare and non-present pattern mining problems will be instantiated from the generalized framework by the quadruplet $(2^I, 2^I, \subseteq, \subseteq)$ and by the pattern families $F_{\delta_1,\delta_2=|D|+1}^D$, $F_{\delta_1=1,\delta_2}^D$, $F_{\delta_1=0,\delta_2=1}^D$.

## 5. RARE AND NON-PRESENT ITEM-SET MINING

Our claim is that all non-frequent item-sets (rare and non-present item-sets) contained in a transaction database can be efficiently mined using an Apriori-like approach by traversing the item-set lattice in a reverse way compared to the traversal performed in the classical Apriori approach. The mining approach benefits of the same advantages as Apriori does. In other words, the backward traversal method is endowed with a property that leads to prune out potentially frequent item-sets in the mining process. Starting from this assumption, we show how the approach is effective and practical for finding all rare and non-present item-sets using an example and then we present the theoretical foundations of the approach. Hereafter, we give the intuition behind our approach to mine rare and non-present item-sets using an Apriori-like approach. Let *D* be the transaction database presented in *Table 1*, and *I* the set of items contained in a transaction of the database *D* and |*I*| the cardinality of *I*. Our goal is to find the set of item-sets that are present in at most two transactions or not present at all. The number of times an item-set occurs in the database is called the item-set support. In our case, the non-inclusive maximum support threshold is set to 3 in the case of rare item-sets, and 1 in the case of non-present item-sets.

### 5.1. Apriori Approach

To find rare and non-present item-sets of size *k-1*, also denoted, respectively, (*k-1*)-rare and (*k-1*)-non-present item-sets, with support under 3 we start by looking for rare and non-present item-sets with size from |*I* | to *k*. In our example, to find the 1-rare item-sets we have to find the 5-rare item-sets, 4-rare item-sets, 3-rare item-sets and 2-rare item-sets. Similarly, to find the 1-non-present item-sets we have to find the 5-non-present item-sets, 4-non-present item-sets, 3-non-present item-sets and 2-non-present item-sets.

The initialization goes through a two stage process. It begins by generating *<abcde>*, the largest item-set, composed of all items in *I*, this item-set is also called a candidate of size 5 or 5-candidate. As we can notice this item-set is not present in any transaction which means its support is null ($supp_D(<abcde>) = 0$) and by then is considered as a non-present item-set (5-non-present item-set). This item-set forms the seed for the next step. After that, we generate the 4-candidates from the non-present item-set *<abcde>* obtained in the previous stage by removing one item at once from it. That is, when we remove the item "*a*" we obtain the candidate *<bcde>*, and by removing items "*b*", "*c*" and "*d*" we obtain the candidates *<acde>*, *<abde>*, *<abce>* and *<abcd>*, respectively. The support of each of those 4-candidates is either 0 or 1 which is less than the fixed maximum support. Consequently, all the 4-candidates are 4-non-present or 4-rare item-sets.

The initialization process ends after generating the 4-rare and 4-non-present item-sets, and then starts a recursive process to generate the remaining item-sets. The 3-candidates are generated by intersecting the obtained item-sets of size 4 that have in common 3 items. For example, <cdei is generated by intersecting hacdei and *<bcde>*, and *<abc>* is the result of the intersection of the rare item-sets *<abcd>* and *<abce>*. By further combining the 4-rare and 4-non-present item-sets we obtain the remaining 3-candidates: *<bde>*, *<bce>*, *<bcd>*, *<ade>*, *<ace>*, *<acd>*, *<abe>*, *<abd>*. Once again, we check the support of each 3-candidate and keep only rare and non-present ones. The only candidate that is discarded from the sets of 3-rare and 3-non-present item-sets is *<abc>* because its support is equal to 3. Similarly to the previous step, the 3-rare and 3-non-present item-sets are combined in such a way to generate the set of 2-candidates by intersecting item-sets having 2 items in common. For example, the intersection of the 3-rare item-sets <cdei and the 3-non-present item-set *<bde>* is the 2-candidate *<de>*. However, the item-sets *<bc>*, *<ac>*, *<ab>* are not considered at all due to the fact that they are sub-item-sets of the frequent item-set *<abc>*. Indeed, both *<bc>*, *<ac>* and *<ab>* have a support equal to 3.

This principle is also called candidate pruning where the item-sets that are sub-item-sets of a frequent item-set are not considered in any combination operation (see Section 0.5.2). The 2-rare item-sets generated at this stage are *<de>*, *<ce>*, *<cd>*, *<be>*, *<bd>*, *<ae>*, *<ad>*. In the last step, the 1-candidates are generated by intersecting the item-sets obtained in the previous step. Due to the pruning step, the item-sets *<b>*, *<c>* and *<d>* are not considered and we can easily verify that the support of both *<b>*, *<c>* and *<d>* is 4. The only 1-rare item-set is *<e>*. The process can stop for two reasons: (1) either no more candidates can be generated or we reach a level where all rare and non-present item-sets are composed of only one item. In our case, the process stops because no more candidates are generated from the 1-rare item-set *<e>*. This example confirms our initial assumptions and encourages us to explore further the approach and build a novel method for rare and non-present item-set mining on a top of a generalized theoretical framework for item-set mining we propose.

What makes Apriori interesting is the ability to avoid the exploration of the whole item-set space. In fact, a great amount of non-frequent item-sets may be discarded in the mining process and only item-sets that are likely to be frequent are explored. The key idea is that if an item-set is non-frequent so are all its supersets. Such property is called the anti-monotone property.

To prune non-frequent item-sets, Apriori exploits the anti-monotone property on a top-down traversal of the item-set space. Hence, the patterns that contain a non-frequent item-set are pruned out from the mining space and only potential candidates are further considered for support test.

In the case of rare and non-present item-sets a top-down traversal of the item-set space cannot benefit of such property to prune out frequent item-sets. In fact, the top-down traversal of the item-set space is not endowed with an equivalent antimonotone property to prune frequent item-sets. That is, with apriori no conclusion can be made about the frequency of an item-set just by knowing that its subsets are frequent item-sets. However, an interesting property is that all supersets of a frequent item-set are frequent.

Our idea is to exploits the later property to prune frequent item-sets on a bottom-up traversal approach. Indeed, when exploring the item-set space from the bottom to the top, if an item-set is not rare and not a non-present pattern than all its supersets are not rare or non-present and by then are pruned out in the mining process. In this respect, instead of growing item-sets to obtain longer ones such as done in the classical Apriori approach, we do the inverse: *reducing item-set sizes at each step starting from the item-set that contains all the items considered in the data set.*

In the next section we present the factorized principles behind the classical Apriori framework and propose a generalized representation which by means of a specific instantiation lead to formalize the Apriori rare and non-present item-set mining approaches based on the principle enounced below.

## 5.2. Level-wise Traversal of the Pattern Space

An Apriori-based approach to mine interesting patterns has to traverse the pattern space and iteratively generate patterns at a level based on the interesting patterns of the precedent level. To determine a level, a range measure has to be defined.

Usually, such measure is closely related to the generality relationship among patterns and defined as a function from $\Gamma$ to N. In other words, the pattern space is layered and then traversed by moving from a level to another. That is, a level is composed of a set of patterns having the same rank value. When traversing the sliced pattern space according to a rank function, all interesting patterns will be reached (see *Proposition 5.1*). The interesting patterns obtained in each pass are used to generate candidate patterns for the next pass. The support of each candidate is then counted, and the interesting ones are kept. This process continues until no new interesting patterns are found or no new candidates are generated.

**Proposition 5.1** *Given a pattern framework $<\Gamma, \Delta, \subseteq_p, \triangleleft>$, the pattern family $F_{\delta_1,\delta_2}^D$ such that the pattern space $\Gamma$ is endowed with a rank function $\rho$, and a partial order that preserves the following monotone property: a pattern at level y is interesting if all the associated patterns at level x w.r.t. the partial order are interesting, then all patterns in $F_{\delta_1,\delta_2}^D$ can be discovered by traversing the levels defined by $\rho$.*

### 5.3. Pattern Composition

Now, to generate a pattern at a level, a pattern operation that merges two or more interesting patterns of a precedent level is performed (see *Definition 5.1*).

**Definition 5.1** *A pattern operation w.r.t. a pattern family is defined as an n-array relation that associates for a given pattern and a list of arguments another pattern in the same pattern family space. More formally, given the pattern framework $<\Gamma, \Delta, \subseteq_p, \triangleleft>$, the family of patterns $F_{\delta_1,\delta_2}^D$, and a pattern $f$ in $F_{\delta_1,\delta_2}^D$, a pattern operation on $f$ and the list of arguments arg1, arg2, … argn w.r.t. to $F_{\delta_1,\delta_2}^D$, denoted by P($f$, arg1, arg2, … argn) is a pattern that belongs to $F_{\delta_1,\delta_2}^D$. The set of pattern operations will be represented by $O_p$.*

In Def*inition 5.1* a pattern operation involves a list of arguments whose semantics are not presented for the sake of generality. Thus, specific arguments with precise semantics have to be set for each pattern operation. For instance, the arguments can be, but not exhaustively, data related to a pattern or to a pattern component. Based on *Definition 5.1*, we present two kinds of more specific operations: (i) generalization operations (see *Definition 5.2*) and (ii) specialization operations (see *Definition 5.3*). Whereas generalization operations lead to a more general pattern than the input patterns, the specialization operations lead to a pattern that is more specific than the combined ones.

**Definition 4.2** *A pattern specialization operation w.r.t. $F_{\delta_1,\delta_2}^D$ is an element of $O_p$ that associates to a given pattern a more specific one. In other words, if we denote this operation by $P_s$ then we have $P_s(f) \subseteq_p f$. The set of specialization operations is denoted by the symbol $S_p$.*

**Definition 4.3** *A pattern generalization operation w.r.t. $F_{\delta_1,\delta_2}^D$ is a relation from $O_p$ that associates to a pattern a more general one. That is, considering the pattern f 2 $F_{\delta_1,\delta_2}^D$ and $G_{op}$ denotes the generalization operation, then $f \subseteq_p G_{op}(f)$. Generalization operations are represented by the set $\Theta_p$.*

*Finall*y, based on the definitions of pattern specialization and pattern generalization operations, we define two more operations that combine two or more patterns and results a pattern which is whether more general or more specific than the combined ones (see *Definition 5.4* and *Definition 5.5*).

**Definition 5.4** *A combining specialization operation, denoted by $C_{sp}$, is a specialization operation that transforms two or more patterns and the result is a pattern which is more specific than the involved patterns. That is to say, $C_{sp}$ is in $S_p$ and if the patterns $f_1, f_2, ..., f_p$ can be combined w.r.t. $C_{sp}$ than $C_{sp}(f_1, f_2, ..., f_p, \arg_{p+1}, \arg_{p+2}, ..., \arg_{p+n}) \subseteq_p f'$ such that $f' \in \{f_1, f_2, ..., f_p\}$.*

**Definition 5.5** *A combining generalization operation, denoted by $C_{gp}$, is a generalization operation that transforms two or more patterns and the result is a pattern which is more general than the involved patterns. For instance, $C_{gp}$ is in $\Theta_p$ and if the patterns $f_1, f_2, ..., f_p$ can be combined w.r.t. $C_{gp}$ than $\forall f \in \{f_1, f_2, ..., f_p\}$ we have $f \subseteq_p C_{gp}(f_1, f_2, ..., f_p, \arg_{p+1}, \arg_{p+2}, ..., \arg_{p+n})$.*

*For exam*ple, as shown in *Section 2* classical Apriori-based approaches generate a pattern at a level *k* by merging two patterns of level *k-1* that share common *k-2* items. The merge operation is reduced to the set union which is a special case of $C_{gp}$ operations.

### 5.5. Pruning Principle

*It is noteworthy that* the move from a level to another can be optimized. For example, some of those optimizations could be to refine the pattern operations by defining canonical operations and by then reduce the redundancy when generating the patterns or by reducing the amount of the candidates that are generated at each pass of the mining process by pruning out candidates that are likely to be not interesting.

The pruning principle is generally based on an anti-monotone property defined on the top of the pattern space endowed with a partial order (see Definition 5.6).

This property may state that some patterns will certainly not be interesting if we do already know that some patterns related to them are not interesting. By doing so, the set of candidates may be considerably reduced. In fact, instead of taking into account all possible combinations of the interesting patterns at a level by means of the combination operations, only subset of it will be considered and the remaining subset will be the patterns that are related to the non-interesting patterns of the precedent level.

**Definition 5.6** *(Anti-monotone and monotone properties) A property $p$ defined on top of the pattern family $F_{\delta_1,\delta_2}^D$ is monotone if and only if whenever a pattern $f \in F_{\delta_1,\delta_2}^D$ satisfies $p$, so does any pattern that is a successor of $f$ w.r.t. the partial order defined on top of $\Gamma$. The property $p$ is anti-monotone if and only if whenever a pattern does not satisfy $p$, so does any super-pattern of $f$ in $F_{\delta_1,\delta_2}^D$.*

The pruning principle enounced in *Proposition 5.2* leads to the reduction of the patterns space. Indeed, in the traversing process only interesting patterns discovered at a level that yield potentially interesting patterns at the next level are considered.

**Proposition 5.2** *Given a pattern framework $<\Gamma, \Delta, \subseteq_p, \triangleleft>$, the pattern family $F_{\delta_1,\delta_2}^D$ and the binary relation R a partial order on $\Gamma$ elements (R is reflexive, anti-symmetric and transitive), then it is possible to prune out part of the pattern space composed of non-interesting patterns during the mining process if it exists an anti-monotone property on $\Gamma$ w.r.t. R.*

To sum up, the generalized Apriori framework is defined by the following elements:

- **Pattern framework**: $<\Gamma, \Delta, \subseteq_p, \triangleleft>$ that define the pattern space, the data set space, the generality and instantiation relationships;
- **Pattern family**: $<D, \delta_1, \delta_2>$ representing the category of patterns to be mined;
- **Rank measure**: a measure $\rho$ to divide the pattern space into levels;
- **Monotone property on the interestingness of patterns:** to ensure the completeness of the process of interesting pattern discovery;
- **Anti-monotone property:** necessary to prune out potentially non interesting patterns;
- **Pattern operations:** to move from a level to another by combining interesting patterns of the first level.

### 5.5. Instantiation of the Generic Apriori Framework for Rare and Non-present Item-set Mining

The instantiation of the generalized Apriori framework let us obtain rare and non-present item-set mining frameworks as follows. The pattern framework we consider is constituted of the quadruplet $(2^I, 2^I, \subseteq, \subseteq)$ where $I$ is a set of items present in the data transactions, both the item-set and data languages are limited to item-sets, and the generalization relation among item-sets is reduced to the set inclusion relationship, so is the instantiation relation among item-sets and data set transactions.

The second component is the item-set family $F_{0,\delta}^D$ where $D$ represents the transaction database and $\sigma$ the maximum support threshold. The rank of an item-set is its cardinality and the item-set space is explored from the higher-rank level to the lower-rank level according to the monotone property: *all supersets of a rare or non-present item-set are rare or non-present*. It lets us say that if an item-set of level $k$ is rare or non-present, so does all its supersets at level $k+1$ which guaranties to reach all rare and non-present item-sets in the traversal process. Furthermore, an item-set of level $k$ is obtained by merging two rare or non-present item-sets of level $k+1$ and the pruning principle given hereafter is used to check that the remaining subsets are also rare or non-present. The operator used to combine the item-sets consists of the set intersection operator (see *Definition 5.7*). For the pruning principle, we use the following anti-monotone property: *all subsets of a frequent item-set are frequent*.

**Definition 5.7** *Let $D$ be the data set, $\sigma$ an integer, and $F_{0,\delta}^D$ the item-set family. An intersection-based generalization operation, denoted by $C_{gp}^i$, is a combining generalization operation that takes two item-sets having $k+1$ items as parameters and generates an item-set that is the intersection of those two item-sets only if they share $k$ items. For instance, $C_{gp}^i$ is a $C_{gp}$ and if $f_1, f_2$ two item-sets from $F_{0,\sigma}^D$ then $C_{gp}^i(f_1, f_2) = f_1 \cap f_2$ only if $|f_1| = |f_2|$ and $f_1, f_2$ share $|f_1|-1$ items. As a result $C_{gp}^i(f_1, f_2) \subset f_1$ and $C_{gp}^i(f_1, f_2) \subset f_2$.*

Once the Apriori-based framework is instantiated, we build an Apriori algorithm, called ARANIM for *Apriori Rare and Non-Present Item-set Mining* to mine rare and non-present item-sets (see *Figure 2*). More specifically, ARANIM performs a level-wise descent in $<\Gamma, \subseteq_p>$. The starting point is the computation of the unique N-rare or N-non-present candidate and the (*N -1*)-rare and or (*N -1*)-non-present item-sets whereby candidates comprise all item-sets having $N - 1$ items (*Line 9 to 11*). Then, the *candTest()* procedure (see *Figure 3*) computes the International Journal on Soft Computing, Artificial Intelligence and Applications (IJSCAI), Vol.1, No.1, August 2012support of each candidate and returns only item-sets having support less than the fixed maximum support (maxsupp = $\sigma$) (*Line 10 to 12*).

For each of the subsequent levels *k*, the candidates are generated by combining the interesting (rare or non-present) (*k+1*)-patterns of the precedent level (*Line 16*). In the prune step, a candidate $f_1 \cap f_2$ is inserted only if all its (*k+1*)-supersets (supersets of cardinality *k+1*) occur in $F_{0,\sigma}^{k+1}$ (*Line 17*). Then, the support of the $F_c^{k+1}$ candidates is counted (*Line 19*). All candidates that turn out to be rare or non-present are inserted to $F_{0,\sigma}^D$.

The process stops whether the set of generated candidates is empty ($F_c^k = \phi$) or rare item-set is generated at the level *k*. The example presented in *Section 5.1* follows the ARANIM process when looking for rare and non-present item-sets on the database of transactions presented in Table 1. It is noteworthy that the item-sets mined are marked in order to have the ability to filter the results and selectively get only rare or non-present item-sets. Also, ARANIM may be used to mine only non-present patterns. In fact, when fixing the maximum threshold to 1, we will only get non-present item-sets.

```
Algorithm 1 ARANIM
 1: Input:
 2:   D, σ;                                              ▷ Set of transactions, support threshold
 3:   C^D_{0,σ};                                         ▷ itemsets contained in D
 4: Output:
 5:   F^D_{0,σ};                                         ▷ Set of rare and non-present itemsets
 6: Initialization:
 7:   I ← set of items contained in the transactions of D;
 8:   N ← |I|;
 9:   F^N_c ← I;                                         ▷ N-candidate
10:   F^N_{0,σ} ← CANDTEST(F^N_c, D);                    ▷ N-rare and N-non-present
11:   F^{N-1}_c ← {f − {i} : f ∈ F^N_{0,σ} ∧ i ∈ I};     ▷ (N − 1)-candidates
12:   F^{N-1}_{0,σ} ← CANDTEST(F^{N-1}_c, D, σ);         ▷ (N − 1)-rare and (N − 1)-non-present
      itemsets
13: Method:
14: for (k = N − 2; F^{k+1}_{0,σ} ≠ ∅; k − −) do
15:     F^k_c ← ∅;
16:     F^k_c ← {f ∩ f' : f, f' ∈ F^{k+1}_{0,σ} ∧ |f ∩ f'| = k};   ▷ (k)-candidates
17:     F^k_c ← F^k_c − {f : f ∈ F^k_c ∧ ∃f' ∈ (C^D_{0,σ} − F^{k+1}_{0,σ}) ∧ |f'| = k + 1 ∧ f ⊂ f'};
18:                                                      ▷ prune out the frequent (k)-candidates
19:     F^k_{0,σ} ← CANDTEST(F^k_c, D, σ);               ▷ k-rare and k-non-present itemsets
20:     F^D_{0,σ} ← F^D_{0,σ} ∪ F^k_{0,σ};
21: end for
22: return F^D_{0,σ};
```

Figure 2: Pseudo code of ARANIM algorithm.

```
Algorithm 2 Candidate test
 1: procedure CANDTEST(F : set of itemsets, D: set of transactions, σ : maximum
    support threshold)
 2:     F_temp ← ∅;
 3:     for all i ∈ F do
 4:         count ← 0;
 5:         for all t ∈ D do
 6:             if i ⊆ t then
 7:                 count + +;
 8:             end if
 9:         end for
10:         if count < σ then
11:             if count! = 0 then
12:                 i.isRare = true;
13:             end if
14:             F_temp ← F_temp ∪ {i};
15:         end if
16:     end for
17:     return F_temp;
18: end procedure
```

*Figure 3: Pseudo code of CandidateTest procedure.*

## 6. EVALUATION WITH THE DISCOVERY OF SUSPICIOUS USAGE DETECTION

Securing Web applications involves encrypting sensitive traffic and data, restricting access to defined parts of the application and preventing misuse of the application.

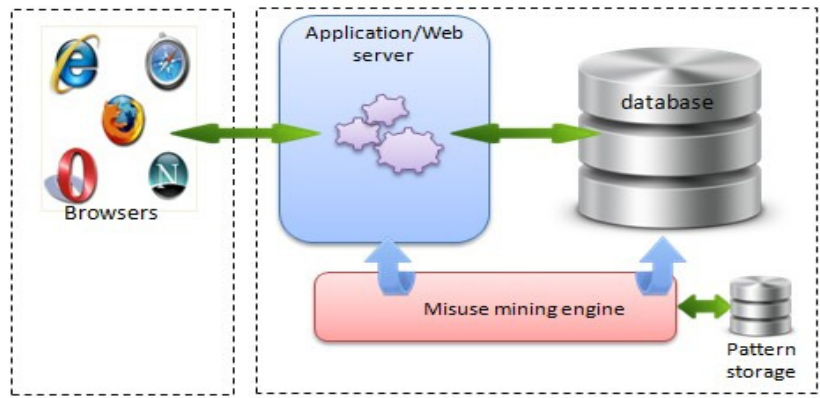

Figure 4: General overview of a Misuse Pattern Discovery system based on Rare Patterns.

In this section, we propose an approach based on rare patterns to detect suspicious uses and behaviors in the context of a Web application. The mining system—called RPMSUD for *Rare Pattern Mining for Suspicious Use Detection*— is mainly composed of an engine that unobtrusively analysis the usage data of a web application and detects suspicious use of a running web application (see *Figure 4*). Our assumption is that a web application usage is dominated with repetitive access/requests and that rare access patterns are source of potential abnormal usage that may be related to security issues.

The algorithm behind RPMSUD system (see *Figure 5*) is mainly composed of an event queue processing process that periodically analyses the events that occurs in a given period of time. Here we call those periods *mining cycles*. We consider a usage pattern suspicious if it is detected during the different mining cycles. The user has to determine the duration of a cycle and the number of cycles to consider before issuing an alert. After each cycle the queue is emptied. The number of cycles multiplied with the duration of a cycle is called the mining window. After each mining window, the set of mined patterns is saved and then emptied.

```
Algorithm 3 RPMSUD
 1: Input:
 2:   Q, σ;                    ▷ events queue, maximum support threshold
 3:   E;                                                  ▷ set of events
 4:   W;                                      ▷ mining window in milliseconds
 5:   cycles, duration;  ▷ number of (mining) cycles in a mining window, duration of
      a cycle in milliseconds (duration×cycles ≤ W)
 6:   AlertAction();           ▷ the action to execute if an alert is triggered

 7: Initialization:
 8:   F ← ∅;                              ▷ set of discovered rare patterns

 9: Method:
10:   for (i ∈ [1, cycles]; wait ← duration) do
11:       F+ = ARANIM(Q, σ);
12:       Q ← ∅;
13:   end for

14:   for (f ∈ F) do
15:       if f.count ≥ cycles then
16:           AlertAction();
17:       end if
18:   end for

19:   store(F);
20:   F ← ∅;
21:   Q ← ∅;
```

Figure 5: Pseudo code of RPMSUD algorithm.

## 7. CONCLUSION AND FUTURE WORK

In this paper we have enhanced the frameworks presented in [1]. Those frameworks consist on:
- A structural framework to represent the different categories of patterns based on the frequency constraint which by means of an instantiation process leads to the representation of frequent, rare and non-present pattern mining problems.
- A mining framework composed of an abstract model that factorizes the classical Apriori approach.

Then, an Apriori method is instantiated to tackle the specific problems of rare and non-present item-set mining. We also presented an overview of an approach—called RPMSUD—that uses rare patterns to detect abnormal usage in web applications. The main advantage of the RPMSUD is that it does not require the construction of a behavioural model of users, and it does not require the extraction of user profiles. Thus, it is flexible and may detect suspicious behaviours not seen before. It is noteworthy that this approach is intended to complete and not compete or replace existing security solutions such as access control, cryptography and intrusion detection systems.

As we are in the initial stages of this research, much remains to be done including—but not limited to—the following tasks:

- Deal with the performance of the mining algorithms by adapting non-Apriori techniques such as Eclat [24], FP-Growth [11] and TM [19];
- Develop an incremental algorithm to mine rare patterns;
- Study how the combination of frequent and rare patterns may enhance security of computer systems in general and web application in particular;
- Accuracy and performance comparison with existing approaches.

## REFERENCES

[1] Adda, M., Wu, L., & Feng, Y. (2007). Rare item-set mining. *Proceedings of the Sixth International Conference on Machine Learning and Applications* (pp. 73–80). Washington, DC, USA: IEEE Computer Society.

[2] Agrawal, R., & Srikant, R. (1994). Fast algorithms for mining association rules in Large Databases. *Proceedings of the 20th International Conference on Very Large Data Bases* (pp. 487–499). San Francisco, CA, USA: Morgan Kaufmann Publishers Inc.

[3] Agrawal, R., Imieliński, T., & Swami, A. (1993). Mining association rules between sets of items in large databases. *Proceedings of the 1993 ACM SIGMOD international conference on Management of data* (pp. 207–216). Washington, D.C., United States: ACM.

[4] Amor, N., Benferhat, S., & Elouedi, Z. (2004). Naive Bayes vs decision trees in intrusion detection systems. *ACM Symposium on Applied Computing* (pp. 420–424). New York: ACM Press.

[5] Barbara, D., Couto, J., Jajodia, S., & Wu, N. (2001). Adam: a testbed for exploring the use of data mining in intrusion detection. *Special Section on Data Mining for Intrusion Detection and Threat Analysis* , 15–24.

[6] Chan, P., Mahoney, M., & Arshad, M. (2003). *A machine learning approach to anomaly detection.* Florida Institute of Technology.

[7] Denning, D. (1987). An intrusion-detection model. *IEEE Transactions on Software Engineering* , 222–232.

[8] DongJun, Z., Guojun, M., & Xindong, W. (2008). An Intrusion Detection Model Based on Mining Data Streams. *Proceedings of The 2008 International Conference on Data Mining, DMIN* (pp. 398–403). Las Vegas: CSREA Press.

[9] Gou, M., Norihiro, S., & Ryuichi, Y. (2002). A framework for dynamic evidence based medicine using data mining. *Proceedings of the 15th IEEE Symposium on Computer-Based Medical Systems* (p. 117). Maribor, Slovenia: IEEE.

[10] Haglin, D. J., & Manning, A. M. (2007). On minimal infrequent item-set mining. *Proceedings of the 2007 International Conference on Data Mining, DMIN 2007* (pp. 141–147). Las Vegas, Nevada, USA: CSREA Press.

[11] Han, J., Pei, J., & Yin, Y. (2000). Mining frequent patterns without candidate generation. *In Proceedings of the 2000 ACM SIGMOD international conference on Management of data* (pp. 1–12). New York, NY, USA: ACM Press.

[12]     Iwanuma, K., Takano, Y., & Nabeshima, H. (2004). On anti-monotone frequency measures for extracting sequential patterns from a single very-long data sequence. *IEEE Conference on Cybernetics and Intelligent Systems*, (pp. 213–217).

[13]     Kumar, S., & Spafford, E. (1994). *An application of pattern matching in intrusion detection.* Technical Report 94–013, Department of Computer Sciences, Purdue University.

[14]     Lane, T., & Brodley, C. (1997). An application of machine learning to anomaly detection. *Proceedings of the 20th NIST-NCSC National Information Systems Security Conference*, (pp. 366–380).

[15]     Lee, W. (1998). Data mining approaches for intrusion detection. *In Proceedings of the Seventh USENIX Security Symposium* (pp. 6–6). San Antonio, Texas: USENIX Association Berkeley, CA, USA.

[16]     Mansfield, G., Ohta, K., Takei, Y., Kato, N., & Nemoto, Y. (2000). Towards trapping wily intruders in the large. *Computer Networks* , 659–670.

[17]     Pasquier, N., Bastide, Y., Taouil, R., & Lakhal, L. (1999). Efficient Mining Of Association Rules Using Closed Item-set Lattices. *Information Systems*, 25–46.

[18]     Prati, R. C., Monard, M. C., André, C. P., & de Carvalho, L. F. (2004). A method for refining knowledge rules using exceptions, SADIO. *Electronic Journal of Informatics and Operations Research* , 53–65.

[19]     Song, M., & Sanguthevar, R. (2006). A transaction mapping algorithm for frequent item-sets mining. *IEEE Transactions on Knowledge and Data Engineering* , 472–481.

[20]     Srikant, R., & Agrawal, R. (1997). Mining Generalized Association Rules. *Future Generation Computer Systems* , 161–180.

[21]     Szathmary, L., Maumus, S., Petronin, P., Toussaint, Y., & Napoli, A. (2006). Vers l'extraction de motifs rares. *EGC: Extraction et Gestion de Connaissances*, (pp. 499–510).

[22]     Tan, K., Killourhy, K., & Maxion, R. (2002). Undermining an Anomaly-Based Intrusion Detection System Using Common Exploits. *Recent Advances in Intrusion Detection* , 54–73.

[23]     Tiwari, A., Gupta, R., & Agrawal, D. (2010). A survey on frequent pattern mining: Current status and challenging issues. *Inform. Technol* , 1278–1293.

[24]     Zaki, M. J., Parthasarathy, S., Ogihara, M., & Li, W. (1997). New algorithms for fast discovery of association rules. *In Proceedings of the 3rd International Conference on Knowledge Discovery and Data Mining*, (pp. 283–296).

[25]     Zhan, J., & Leshan Teachers Coll., L. (2008). Intrusion Detection System Based on Data Mining. *First International Workshop on Knowledge Discovery and Data Mining, 2008. WKDD 2008*, (pp. 23–24). Adelaide, SA.

**Authors**

**Mehdi Adda,** Professor of computer science at the University of Quebec in Rimouski, Rimouski, Canada. His principal research interests lie in the fields of software and web engineering, data mining and knowledge discovery, aspect oriented programming and distributed computing, web personalization and recommendation. Contact him at adda@ieee.org or mehdi_adda@uqar.qc.ca.

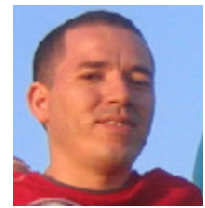

**Lei Wu,** Assistant Professor of software engineering at University of Houston-Clear Lake, Houston, U.S.A. His major research interests include software engineering with artificial intelligence, secure service-oriented architectures, software for robotics and embedded system intelligence, game software development, and pervasive computing. He can be reached at wul@uhcl.edu

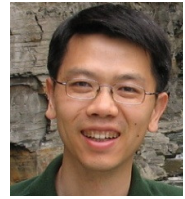

**Sharon White,** Associate Professor of software engineering at University of Houston-Clear Lake, Houston, U.S.A. Her research interests includeinclude domain specification languages, architecture design languages, and software architecture. She can be reached at whites@uhcl.edu

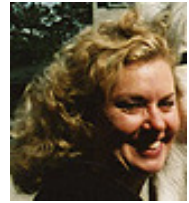

**Yi Feng,** Associate Professor of computer science, Algoma University, Sault Ste. Marie, Canada. Her major research interests include formal verification, software engineering, signal processing, and system description languages. She can be reached at feng@algomau.ca

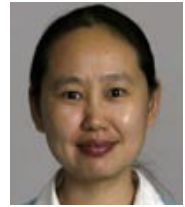